\documentstyle[amsfonts,amsmath,epsfig,12pt]{article}

\newcommand {\slsh} [1] {\not{\hbox{\kern-2pt${#1}$}}}

\newcommand{\drawsquare}[2]{\hbox{%
\rule{#2pt}{#1pt}\hskip-#2pt
\rule{#1pt}{#2pt}\hskip-#1pt
\rule[#1pt]{#1pt}{#2pt}}\rule[#1pt]{#2pt}{#2pt}\hskip-#2pt
\rule{#2pt}{#1pt}}

\newcommand{\Yfund}{\raisebox{-.5pt}{\drawsquare{6.5}{0.4}}}
\newcommand{\Yasymm}{\raisebox
{-3.5pt}{\drawsquare{6.5}{0.4}}\hskip-6.9pt%
                      \raisebox{3pt}{\drawsquare{6.5}{0.4}}%
                     }
\newcommand{\Ysymm}{\Yfund\hskip-0.4pt%
                     \Yfund}

\newcommand\T{\rule{0pt}{2.5ex}}

\newcommand\B{\rule[-1.7ex]{0pt}{0pt}}

\def\drawbox#1#2{\hrule height#2pt
         \hbox{\vrule width#2pt height#1pt \kern#1pt
               \vrule width#2pt}
               \hrule height#2pt}

\def\Asym#1#2{\vcenter{\vbox{\drawbox{#1}{#2}
               \kern-#2pt       
               \drawbox{#1}{#2}}}}

\def\bdot{\huge{\textbf{.}}}

\newcommand {\beq} {\begin{equation}}
\newcommand {\eeq} {\end{equation}}
  \newcommand {\ber}{\begin{eqnarray*}}
  \newcommand {\eer} {\end{eqnarray*}}
\newcommand {\bea}{\begin{eqnarray}}
  \newcommand {\eea} {\end{eqnarray}}

\newcommand{\Dslash}{\,{\raise.15ex\hbox{/}\mkern-12mu D}}

\begin{document}


\begin{titlepage}

\begin{center}
\vspace{1in}
\large{\bf A Note on Seiberg Duality}\\
\large{\bf and Chiral Symmetry Breaking}\\
\vspace{0.4in}
\large{Adi Armoni}\\
\small{\texttt{a.armoni@swan.ac.uk}}\\
\vspace{0.2in}
\emph{Department of Physics, Swansea University}\\ 
\emph{Singleton Park, Swansea, SA2 8PP, UK}\\
\vspace{0.3in}
\end{center}

\abstract{Following arXiv:1310.2027 and arXiv:0801.0762, we consider
 a non-supersymmetric Seiberg duality between electric and magnetic ``orientifold field theories''. These theories live on brane configurations of type 0' string theory. In the electric theory side the scalars acquire a mass and decouple, resulting in an $SU(N_c)$ gauge theory coupled to $N_f$ massless quarks and an additional massless fermion that transforms in the two-index antisymmetric representation. In the magnetic theory side there exists a fundamental meson field that develops a Coleman-Weinberg potential. At the one-loop approximation the potential admits a minimum with chiral symmetry breaking of the form 
$SU(N_f)_L\times SU(N_f)_R \rightarrow SU(N_f)_V$ and an additional breaking of an axial $U(1)$ symmetry. The resulting theory admits a spectrum whose massless degrees of freedom are $N_f^2$ Nambu-Goldstone bosons.}

\end{titlepage}

\section{Introduction}

It is widely believed that QCD admits confinement and chiral symmetry breaking due to the formation of the chiral quark condensate $\langle \bar q^i q_j \rangle= \delta ^i _j \Lambda ^3$.
Starting with $SU(N_c)$ gauge theory coupled to $N_f$ massless flavors, the IR degrees of freedom of the theory are expected to be the Nambu-Goldstone bosons that correspond to chiral symmetry breaking of the form
\beq
SU(N_f)_L\times SU(N_f)_R \rightarrow SU(N_f)_V \,, \label{csb}
\eeq
namely $N_f^2-1$ massless mesons. However, even after four decades of intensive studies of QCD, the mechanism behind chiral symmetry breaking is not fully understood.

A promising approach, which led to insights about the IR degrees of freedom of strongly coupled gauge theories, is Seiberg duality \cite{Seiberg:1994pq}. Unfortunately the vast majority of the literature is restricted to super QCD.

In this note we would like to use Seiberg duality between a pair of two non-supersymmetric gauge theories in order to obtain new insights about the IR degrees of freedom of real QCD.

In ref.\cite{Armoni:2008gg} a Seiberg duality between a pair of two non-supersymmetric gauge theories was proposed. The results of \cite{Armoni:2008gg} were mainly about the planar limit, where the theory acquires supersymmetry in a well defined sector \cite{Armoni:2003gp}. The purpose of this note is to focus on the dynamics of the theory at finite-$N_c$, with $N_f$ such that the theory is not conformal in the IR. Our derivations are closely related to those presented recently in \cite{Sugimoto:2012rt,Armoni:2013ika}, see also \cite{Hook:2013vza}.

The main outcome of the present non-supersymmetric duality is that a QCD-like theory has an IR description in terms of mesons. In fact, contrary to the supersymmetric case, the only massless degrees of freedom are the Nambu-Goldstone mesons that correspond to chiral symmetry breaking of the form \eqref{csb}, plus an additional Nambu-Goldstone boson that corresponds to the breaking of an axial $U(1)$ symmetry.

The paper is organized as follows: in section 2 we present the dual electric and magnetic theories at the classical level. In section 3 we discuss how quantum effects modify the classical picture. Finally in section 4 we summarize our results.

\section{Electric and Magnetic theories}

We consider a pair of electric and magnetic theories that live on brane
configurations of non-critical type 0' string theory \cite{Sagnotti:1995ga,Sagnotti:1996qj}. The matter content of the electric-magnetic pair and the duality between them was proposed in \cite{Armoni:2008gg}.

Both the electric and magnetic theories admit an anomaly free $SU(N_f)_L\times SU(N_f)_R\times U(1)_R$ global symmetry. R-symmetry is a name for an anomaly free axial symmetry, borrowed from the supersymmetric jargon. The matter content of both theories is listed in table \eqref{tableelectric} and table \eqref{tablemagnetic} below.

The electric theory is similar to QCD. In addition to the quarks of the theory, there exists a fermion that transforms in the antisymmetric representation of $U(N_c)$ (a ``gluino''). In addition there are scalar quarks (``squarks''). As we shall see, these scalar fields will acquire a mass due to quantum corrections and hence will decouple from the low energy theory.

\begin{table}[!ht]
\begin{center}
\begin{tabular}{|c|c|ccc|}
\hline
		\multicolumn{5}{|c|} {Electric Theory} \\
\hline \hline
               & $\operatorname U(N_c)$ & $\operatorname{SU}(N_f)_L$ & $\operatorname{SU}(N_f)_R$ & $\operatorname U(1)_R$ \\
\hline
$A_\mu$            & adjoint & \bdot & \bdot & 0 \\
                   & $N_c^2$ & & & \\
\hline
$\lambda$ & $\T\Yasymm\B$    &  \bdot  &  \bdot  &  1 \\
                   & $\tfrac{N_c(N_c-1)}{2}\B$  & & &  \\
                   \hline
$\tilde\lambda$ 
	& $\T\overline\Yasymm\B$  &  \bdot  &  \bdot  &  1 \\
                   & $\tfrac{N_c(N_c-1)}{2}\B$  & & &  \\
\hline\hline
$\Phi$   & $\overline\Yfund$ & $\Yfund$ & \bdot & $\T\frac{N_f-N_c+2}{N_f}$ \\
&   $\bar{N_c}$  &   $N_f$ &   & \\
\hline
$\Psi$ & $\Yfund$ & $\Yfund$ & \bdot & $\T\frac{-N_c+2}{N_f}$ \\
  &   $N_c$  &   $N_f$ &  & \\
\hline\hline
$\tilde \Phi$	& $\Yfund$ & \bdot & $\overline\Yfund$  & $\T\frac{N_f-N_c+2}{N_f}$ \\
&   $N_c$  &  & $\bar{N_f}$ & \\
\hline
$\tilde\Psi$	& $\overline \Yfund$ & \bdot & $\overline \Yfund$  & $\T\frac{-N_c+2}{N_f}$  \\
&   $\bar{N_c}$  & & $\bar{N_f}$ &\\
\hline
\end{tabular}
\caption{\it The matter content of the electric theory.}
\label{tableelectric}
\end{center}
\end{table}

The magnetic theory is a $U(\tilde N_c)$ gauge theory, with $\tilde N_c = N_f - N_c +4$. Apart from fields that are charged under the gauge group, it also contains a meson field that transforms in the bifundamental of $SU(N_f)_L \times SU(N_f)_R$. In addition the theory contains ``mesino'' fields that transform in either the symmetric of $SU(N_f)_L$ or the conjugate symmetric of $SU(N_f)_R$.

\begin{table}[!ht]
\begin{center}
\scalebox{.93}[1]{
\begin{tabular}{|c|c|ccc|}
\hline
		\multicolumn{5}{|c|} {Magnetic Theory $\enspace(\tilde{N_c}=N_f-N_c+4)$} \\
\hline\hline
               & $\T\operatorname U(\tilde{N_c})$ & $\operatorname{SU}(N_f)_L$ & $\operatorname{SU}(N_f)_R$ & $\operatorname U(1)_R$  \\
\hline \hline
$A_\mu$            & adjoint & \bdot & \bdot & 0 \\
                   & $\tilde{N_c^2}$ & & & \\
\hline
$\lambda         $ & $\T\Yasymm\B$           &  \bdot  &  \bdot  &  1 \\
 & $\tfrac{\tilde N_c(\tilde N_c-1)}{2}\B$  & & & \\
\hline
$\tilde\lambda $ & $\T\overline\Yasymm\B$  &  \bdot  &  \bdot  &  1 \\
 & $\tfrac{\tilde N_c(\tilde N_c-1)}{2} \B$  & & & \\
\hline\hline
$\phi$ & $\Yfund$ & $\overline\Yfund$ & \bdot & $\T\frac{N_c-2}{N_f}$\\
 &   $\tilde{N_c}$  &   $\bar{N_f}$ &   & \\
\hline
$\psi$ & $\overline \Yfund$ & $\overline \Yfund$ & \bdot & $\T\frac{N_c-N_f-2}{N_f}$ \\
&   $\bar{\tilde{N_c}}$  &   $\bar{N_f}$ &   & \\
\hline\hline
$\tilde \phi$ & $\overline\Yfund$ & \bdot & $\Yfund$  & $\T\frac{N_c-2}{N_f}$ \\
& $\bar{\tilde{N_c}}$  &  & $N_f$ &  \\
\hline
$\tilde\psi$ &   $\Yfund$ & \bdot & $\Yfund$  & \\
& $\tilde{N_c}$  &  & $N_f$ & \\
\hline\hline
M &   \bdot & $\Yfund$ & $\overline \Yfund$ & $\T\frac{2N_f-2N_c+4}{N_f}$ \\
  & & $N_f$ & $\bar{N_f}$ & \\
\hline
$\chi$ &   \bdot & $\Ysymm$ & \bdot & $\T\frac{N_f-2N_c+4}{N_f}$ \\
   &   & $\tfrac{N_f(N_f+1)}{2}$ & & \\
\hline 
$\tilde\chi$ & \bdot & \bdot & $\overline \Ysymm$ & $\T\frac{N_f-2N_c+4}{N_f}$ \\
 &   & & $\T\tfrac{N_f(N_f+1)}{2}$ & \\
\hline
\end{tabular}}
\caption{\it The matter content of the magnetic theory.}
\label{tablemagnetic}
\end{center}
\end{table}

The dual pair admits a realization in non-critical type 0' string theory.
The duality between the two theories has been derived by using a worldsheet argument \cite{Murthy:2006xt,Ashok:2007sf}. It can also be derived by using the ``swapping branes'' argument of \cite{Elitzur:1997fh}.

We would also like to mention that the duality is supported by 't Hooft matching of global anomalies \cite{Armoni:2008gg}
\bea
& SU(N_f)^3 & N_cd^3(\Yfund) \, ,  \\
& SU(N_f)^2 U(1)_R & \frac{-N_c^2+2N_c}{N_f} d^2(\Yfund) \, , \\
& U(1)^3_R & N_c (N_c-1-2\frac{(N_c-2)^3}{N_f^2} ) \, , \\
& U(1)_R & -N_c^2+3N_c \,.
\eea

In the Veneziano large-$N$ limit the electric theory \eqref{tableelectric} becomes equivalent to the electric $U(N_c)$ SQCD and similarly the magnetic theory \eqref{tablemagnetic} becomes equivalent to the magnetic $U(\tilde N_c)$ SQCD theory \cite{Armoni:2003gp,Armoni:2008gg}. Hence the large-$N_c$ Seiberg duality between the theories \eqref{tableelectric} and \eqref{tablemagnetic} is straightforward.

Encouraged by the matching of the global anomalies at any $N_c$ and $N_f$ we propose that the duality holds not only at infinite $N_c$ but at finite-$N_c$ as well. As we shall see, quantum effects will lead to the decoupling of certain fields, resulting in a duality between a QCD-like electric theory and a magnetic theory of massless Nambu-Goldstone mesons.

\section{Quantum effects}

The electric and the magnetic theories are both non-supersymmetric. Therefore scalar field will admit a non-trivial potential, induced by quantum loops. 

\subsection{Electric theory}

The electric theory contains squark fields $\Phi$ and $\tilde \Phi$. The squark field couples to the gluon and to the antisymmetric gluino. Let us consider the one-loop correction the the squark mass. There are $N_c^2$ gluons and $N_c^2-N_c$ gluinos. Therefore, unlike the supersymmetric case, the imbalance between the bosonic and fermionic degrees of freedom that run in the loop lead to a positive mass square for the squark fields
\beq
 M^2_\Phi = M^2 _{\tilde \Phi} = g_e^2 \Lambda ^2 \, ,
\eeq
where $g_e$ is the electric gauge coupling and $\Lambda$ is the UV cut-off of the theory. We therefore expect that the squarks will decouple from the low-energy dynamics.

In addition, similarly to $U(N_c)$ SQCD, the coupling of the $U(1)$ gauge boson (the ``photon'') flows to a zero value in the IR, due to the coupling of the photon to the quarks. 

Thus the low-energy electric theory becomes an $SU(N_c)$ gauge theory coupled to $N_f$ massless quarks and an additional massless Dirac fermion that transforms in the antisymmetric representation. Such a theory is expected to confine and to admit chiral symmetry breaking. According to both Coleman-Witten argument \cite{Coleman:1980mx} and Vafa-Witten theorem \cite{Vafa:1983tf}, the pattern of the chiral symmetry breaking is given by eq.\eqref{csb}.

\subsection{Magnetic Theory}

Similarly to the electric theory, the magnetic theory contains scalar quarks (``squarks''). The squarks couple via a gauge interaction (denoted by $g_m$) to the gluon and to the gluino and via a ``Yukawa'' interaction (denoted by $y$) to the meson and to the mesino. The Yukawa interaction is inherited from the supersymmetric theory, due to the superpotential $W=MQ\tilde Q$. Moreover, for $SU(N_c)$ SQCD, when $N_c < N_f < {3\over 2} N_c$, the couplings $g_m$ and $y$ are related to each other \cite{Oehme:1998yw}
\beq
{g^2_m \over y^2}= {(3{N_f\over N_c}-1)({N_f\over N_c} -1)\over ({N_f\over N_c} -1 -{2\over N_c^2})} \,.
\eeq
We assume (or postulate) that the relation between $g_m$ and $y$ is the same as in the supersymmetric theory, and in particular $g_m > y$.

We now turn to the calculation of the squark mass. The details of the calculation are very similar to those presented in \cite{Armoni:2013ika}. The generated mass at the one-loop level is
\beq
M^2_\phi = M^2_{\tilde \phi}=(g^2 _m - y^2) \Lambda ^2 \, , 
\eeq
hence $M^2_\phi = M^2_{\tilde \phi} = c g^2_m \Lambda^2 >0$ (with $c\sim 1$). We conclude that the magnetic theory contains $N_f$ massless quarks and $N_f$ massive squarks. 

The meson field $M^{\tilde i}_i$ couples to the quarks and the squarks. The mass splitting between them generates a Coleman-Weinberg potential for the meson. By using a chiral transformation it is possible to diagonalize the matrix $M^{\tilde i} _i$
\beq
yM^{\tilde i} _i = {\rm diag} (\lambda _1, \lambda_2, ..., \lambda _{N_f}) \,.
\eeq
The generated one-loop Coleman-Weinberg potential takes the following form 
\bea 
& & 
 V(\{ \lambda_i \})= \label{potential} \\
& &
 \tilde N_c \left ( {\rm tr} \int {d^4 p \over (2\pi)^4} \log (p^2 +y^2 (MM^\dagger) + cg_m^2 \Lambda ^2) - {\rm tr}\int {d^4 p \over (2\pi)^4} \log (p^2 +y^2 (MM^\dagger)) \right ) \nonumber \\
& &
={\tilde N_c \over 4\pi ^2} \sum_{i=1} ^{N_f} \left \{ (\lambda _i ^2 +cg_m^2 \Lambda ^2)^2 \log  (\lambda _i ^2 +cg_m^2 \Lambda ^2) -  (\lambda _i ^2)^2 \log (\lambda _i ^2) \right \} . \nonumber
\eea

The potential \eqref{potential} is plotted in figure \eqref{plot} below.
\begin{figure}[!ht]
\centerline{\includegraphics[width=8cm]{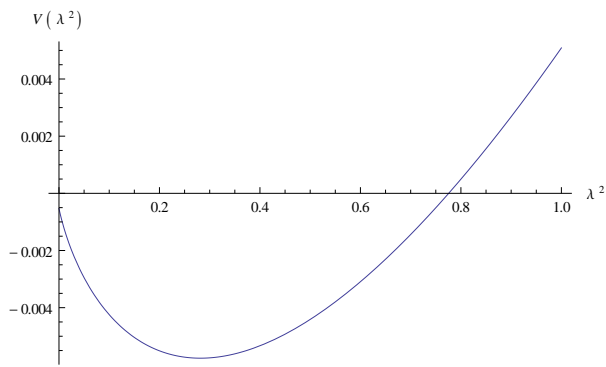}}
\caption{\footnotesize The potential for the eigenvalue of the meson field (divided by the cut-off). In this plot we used $cg_m^2=0.005$.}
\label{plot}
\end{figure}

 The potential admits a unique minimum where all the eigenvalues of $M^{\tilde i} _i$ coincide
\beq
\langle M^{\tilde i} _i \rangle \sim \Lambda \delta ^{\tilde i}_i \, , \label{vev}
\eeq
namely the vacuum configuration breaks $SU(N_f)_L \times SU(N_f)_R$ to $SU(N_f)_V$ as well as $U(1)_R$ completely.

The breaking 
\beq 
SU(N_f)_L \times SU(N_f)_R \times U(1)_R \rightarrow SU(N_f)_V \label{csb2}
\eeq
 results in $N_f^2$ Nambu-Goldstone bosons that correspond to flat directions of the potential \eqref{potential}. 

By using the SQCD relation between the meson field and the electric variables, 
\beq
M^{\tilde i} _i = \tilde \Phi^{\tilde i}_a \Phi^a_i
\eeq
and the equations of motion for the electric squark fields
\bea 
& & \Phi ^a _i \sim \lambda ^{[ab]} \Psi _{bi} \\
& & \tilde \Phi ^{\tilde i}_a \sim \tilde \lambda_{[ab]} \tilde \Psi ^{\tilde ib}
\eea
We conclude that the breaking \eqref{csb2} due to the vacuum expectation value of the meson field \eqref{vev} can be written in the electric variables as a four-quark condensate of the form
\beq 
\langle  \tilde \lambda _{[ab]}\tilde \Psi ^{\tilde i b} \lambda ^{[ac]} \Psi  _{ci} \rangle \sim \Lambda ^6 \delta ^{\tilde i}_i \, .
\eeq
As a consistency check of our derivation we note that the four-quark condensate carries the same $R$-charge as the meson field, $R={2N_f - 2N_c +4 \over N_f}$.

At the minimum of the potential both the quarks and the squarks acquire a (large) mass proportional to $\langle M^{\tilde i} _i \rangle$, due to Yukawa terms. The large mass of the matter fields leads to a decoupling of the massless Nambu-Goldstone mesons from the ``vector multiplet'' (the gluon and the gluino). The $SU(\tilde N_c)$ gauge theory will confine and will acquire a mass gap, due to the formation of massive glueballs. In addition there will be a free $U(1)$ ``photon'', that can be identified with the corresponding photon of the electric theory.

\section{Summary}

In this paper we studied a Seiberg duality between a QCD-like $U(N_c)$ electric theory and a $U(\tilde N_c)$ magnetic theory. The electric theory flows at low energies to an $SU(N_c)$ gauge theory with $N_f$ massless quarks and an additional massless anti-symmetric fermion (plus an additional ``photon''). The theory admits a $SU(N_f)_L \times SU(N_f)_R \times U(1)_R$ global symmetry. The global symmetry may get broken due a formation of quark condensates. 

We analyzed the fate of the global symmetry by using a weakly coupled magnetic theory. We used a one-loop analysis that indicated a breaking pattern of the form \eqref{csb2}.

We must stress that the one-loop analysis cannot be fully trusted. At the minimum of the potential the vacuum expectation value of the meson field is large and the magnetic theory becomes strongly coupled. The one-loop Coleman-Weinberg can only be used to argue that the symmetric point $M^{\tilde i} _i =0$ is unstable. Nevertheless, our result is consistent with both Coleman-Witten \cite{Coleman:1980mx} and Vafa-Witten \cite{Vafa:1983tf}.

It will be interesting to explore other aspects of QCD by using the proposed duality.
\vskip 0.5cm
{\it \bf Acknowledgments.} I wish to thank to professor Graham Shore for a careful reading of the manuscript.


\end{document}